\documentstyle[12pt]{article}
\begin{document}

\thispagestyle{empty}
{}~ \hfill TECHNION-PH-96-8 \\[2mm]
\vspace{0.3cm}

\begin{center}
{\large \bf Weak  Radiative  Decays of Beauty Baryons}\\[5mm]
\vspace{0.5cm}
{\bf Paul Singer and Da-Xin Zhang}\\[5mm]
\vspace{0.20cm}
Department of Physics,
Technion -- Israel Institute of Technology,\\[3mm]
Haifa 32000, Israel\\[5mm]
\vspace{1cm}
{\bf Abstract}
\end{center}

\noindent 
Weak radiative decays of  beauty baryons into strange baryons,
induced by the electroweak penguin,
are estimated by using a quark model approach.
Relations between formfactors
in the semileptonic and in the weak radiative decays
are derived within the heavy quark effective theory.
The partial decay widths are found to be of the order of
 $10^{-15}{\rm MeV}$ for $\Lambda_b\rightarrow\Lambda\gamma$ and  
$\Xi_b\rightarrow\Xi\gamma$
and  of the oder of $10^{-13}{\rm MeV}$ for $\Omega_b\rightarrow\Omega\gamma$.
The $\Omega_b$ radiative decay is thus expected at the sizable branching ratio
of approximately $10^{-4}$.

\vspace{1.2cm}

\noindent
PACS number(s): 12.20.Ds, 12.39Hg, 14.20Mr


\newpage
The investigation of the  electroweak penguin transition
$b\to s\gamma$ is of prime importance,
both as a  test of the standard model\cite{1}
and as a possible window of new physics\cite{2}.
The recent observation of the exclusive process $B\rightarrow K^*\gamma$
 with a branching ratio of $(4.5\pm 1.5\pm 0.9)\times 10^{-5}$\cite{3}
and of the inclusive process $B\rightarrow X_s\gamma$
with a branching ratio of $(2.32\pm 0.57\pm 0.35)\times 10^{-4}$\cite{4}
constitute solid evidence in the support of the interpretation
of these decays in terms of the short-distance
$b\to s\gamma$ transition.

The theoretical treatment of this basic loop process using 
renormalization group equations to treat perturbative QCD 
corrections has been improved during recent years\cite{5},
following the original assertion\cite{6} on the
role of the QCD corrections in increasing the rate of this process
 to bring it into the realm of observability.
The full leading order calculation of the process has been 
completed\cite{7} and the next-to-leading order calculation has also been
partially  performed\cite{8}.
At present, the theoretical uncertainty of the standard model
calculations is of the order of $30\%$\cite{9}
and the experimental errors are even slightly higher\cite{3,4}.

Along with the improvement in the theoretical calculations
by the completion of the next-to-leading order calculation
which should reduce the  theoretical uncertainty,
and the expected increased accuracies of future measurements,
it is of obvious interest to investigate additional physical
processes which are driven by the $b\to s\gamma$ transition.
The weak radiative decays of  beauty baryons
are natural candidates for this task.
Indeed, preliminary estimates on such decays were undertaken recently by
Cheng {\it et al}\cite{10} and by Cheng and Tseng\cite{11} and an
overview of this topic is given in Ref. \cite{12}.

In the present work, the  weak radiative decays
of the beauty baryons are investigated by the use
of  the quark model employed by 
Hussain  {\it et al}\cite{13}
and by K\"orner and Kr\"amer\cite{14} to treat
the semileptonic and the nonleptonic decays
of heavy baryons.
In their model, 
spin interactions between the spectator  and active quarks
are ignored and the $q^2$-dependence of all the formfactors
are taken as pole-like.
This approach is  consistent with the
heavy quark effective theory(HQET), as applied to baryons\cite{15}.

The lowest lying baryons
containing one heavy beauty quark
can be classified into $\bar{3}$, $6$ and $6^*$
under the $SU(3)$ flavor symmetry of the light
quarks\cite{15,16}.
From among these states, only  the spin $1/2$ baryons
$\Lambda_b^0$ , $\Xi_b^{-,0}$ of the antitriplet-baryons
and $\Omega_b$ of the sextet-baryons are expected to decay weakly. 
We consider here the weak radiative exclusive processes which can be induced
by the short-distance $b\to s\gamma$ transition;
these are $\Lambda_b^0\rightarrow\Lambda^0\gamma$, 
$\Xi_b^{-,0}\rightarrow\Xi^{-,0}\gamma$[10-12]
and $\Omega_b^{-}\rightarrow\Omega^{-}\gamma$.
Additional radiative decays like $\Lambda_b\to \Sigma_c\gamma$,
$\Xi_b\to\Xi_c\gamma$, which are caused by weak bremsstrahlung quark
processes, were found to be very rare\cite{10} and will not
be of our concern here.

The basic mechanism for the decays considered here is assumed to be 
the quark level transition $b\rightarrow s\gamma$,
given by the following amplitude[5-9]
\begin{equation}
{\cal A}=
\displaystyle\frac{G_F}{\sqrt{2}}\displaystyle\frac{e}{8\pi^2}c_7^{eff}(\mu)
V_{tb}V_{ts}^*F_{\mu\nu}
\bar{s}\sigma_{\mu\nu}[m_b(1+\gamma_5)+m_s(1-\gamma_5)]b,
\label{heff}
\end{equation}
where $V_{tb}$ and $V_{ts}$ are Cabibbo-Kobayashi-Maskawa matrix elements
and $F_{\mu\nu}$ is the field strength tensor of the photon. 
The coefficient $c_7^{eff}(\mu)$,
which is the combination
of several Wilson coefficients running from $\mu\sim m_t$
to $\mu\sim m_b$, has been calculated[7-9] to be
\begin{equation}
c_7^{eff}(m_b=4.5{\rm GeV})=0.32,
\end{equation}
when one uses   $m_t=174{\rm GeV}$ and $\Lambda_{QCD}=200MeV$.
A different choice for the renormalization point $\mu$
introduces the large uncertainty we mentioned.

In order to calculate the baryonic transitions induced by $b\to s\gamma$,
our basic tool is the heavy quark effective theory which permits
to relate\cite{15,17} in the heavy quark limit the formfactors of the magnetic
transition (1) to those of the semileptonic decays.
Moreover, we treat the decays under consideration here as 
heavy-to-light ($b\to s$) quark transitions.
Throughout the present paper, we neglect corrections
in the mass parameter of the order $1/m_b$.
Needless to say, a more accurate approach should consider
the mass correction as well, especially when the above mentioned
uncertainty in (2) will be reduced.

We turn now to derive the formfactors required here.
For a baryonic transition induced by a $V-A$
current between spin $1/2$ baryons,
one generally has six formfactors.
However, in the limit of the heavy quark mass $m_Q\to \infty$,
using HQET one can express the relevant matrix element in terms of
two independent formfactors only\cite{13,15}.
Thus one has for $\Lambda_b\rightarrow\Lambda$ 
(and a similar expression for $\Xi_b\rightarrow\Xi$),
\begin{equation}
<\Lambda|\bar{s}\gamma_\mu (1-\gamma_5)b|\Lambda_b>=
\bar{u}_{\Lambda}(P_2)
[F_1(q^2)+F_2(q^2)\displaystyle\frac{{\not\! P_1}}{m_{\Lambda_b}}]
\gamma_\mu (1-\gamma_5)
u_{\Lambda_b}(P_1).
\label{lamtypeff}
\end{equation}
For the transition  $\Omega_b\rightarrow\Omega$, 
where we have a heavy spin $1/2$ baryon belonging
to the 6-representation decaying into a spin $3/2$ baryon
belonging to the decuplet, the matrix element
involves six independent formfactors\cite{17}, 
which for the $V-A$ current is
\begin{equation}
\begin{array}{rcl}
<\Omega|\bar{s}\gamma_\mu (1-\gamma_5)b|\Omega_b>&&\\[5mm]
&&\hspace{-4.0cm}
={\bar{u}_{\Omega}}(P_2)_{\alpha}
[g^{\alpha\beta}(C_1+C_2\displaystyle\frac{\not\! P_1}{m_{\Omega_b}})
+\displaystyle\frac{{P_1}^\alpha}{m_{\Omega_b}}
\displaystyle\frac{{P_1}^\beta}{m_{\Omega_b}}
(C_3+C_4\displaystyle\frac{\not\! P_1}{m_{\Omega_b}})\\[5mm]
&&\hspace{-3.5cm}
+\displaystyle\frac{{P_1}^\alpha}{m_{\Omega_b}}\gamma^\beta
(C_5+C_6\displaystyle\frac{\not\! P_1}{m_{\Omega_b}})]
\gamma_\mu (1-\gamma_5)
\sqrt{\displaystyle\frac{1}{3}}(\gamma_\beta+
\displaystyle\frac{{P_1}_\beta}{m_{\Omega_b}})
u_{\Omega_b}(P_1).
\end{array}
\label{ometypeff}
\end{equation}

Relating now by HQET\cite{15} the formfactors in (3)(4) to the  matrix elements
for the radiative decays,
and contracting these with the electromagnetic tensor $F_{\mu\nu}$,
one obtains
\begin{equation}
F^{\mu\nu}<\Lambda|\bar{s}\sigma_{\mu\nu} (1+\gamma_5)b|\Lambda_b>
=F^{\mu\nu}\bar{u}_{\Lambda}(P_2)
[F_1(q^2)+F_2(q^2)\displaystyle\frac{m_{\Lambda}}{m_{\Lambda_b}}]
\sigma_{\mu\nu}(1+\gamma_5)
u_{\Lambda_b}(P_1)
\label{lamtypeff2}
\end{equation}
for the weak radiative decay $\Lambda_b\rightarrow\Lambda\gamma$ 
(or $\Xi_b\rightarrow\Xi\gamma$), and likewise
\begin{equation}
\begin{array}{rcl}
F^{\mu\nu}<\Omega|\bar{s}\sigma_{\mu\nu}
 (1+\gamma_5)b|\Omega_b>
&=&F^{\mu\nu}{\bar{u}_{\Omega}}(P_2)_{\alpha}
[g^{\alpha\beta}(C_1+C_2\displaystyle\frac{m_{\Omega}}{m_{\Omega_b}})
+\displaystyle\frac{{P_1}^\alpha}{m_{\Omega_b}}
\displaystyle\frac{{P_1}^\beta}{m_{\Omega_b}}
(C_3+C_4\displaystyle\frac{m_{\Omega}}{m_{\Omega_b}})\\[5mm]
&&\hspace{-2.0cm}
+\displaystyle\frac{{P_1}^\alpha}{m_{\Omega_b}}\gamma^\beta
(C_5+C_6\displaystyle\frac{m_{\Omega}}{m_{\Omega_b}})]
\sigma_{\mu\nu}(1+\gamma_5)
\sqrt{\displaystyle\frac{1}{3}}(\gamma_\beta+
\displaystyle\frac{{P_1}_\beta}{m_{\Omega_b}})
u_{\Omega_b}(P_1)
\end{array}
\label{ometypeff2}
\end{equation}
for $\Omega_b\rightarrow\Omega\gamma$.
A possible alternative approach which was used before\cite{10,11,18}
is to establish firstly the relations between formfactors in the rest frame
of the initial heavy hadron and then to boost them
to a general Lorentz frame.
However, certain difficulties occur\cite{11} when carrying out this procedure
for the $\Lambda_b\rightarrow\Lambda\gamma$, 
$\Xi_b\rightarrow\Xi\gamma$ decays.

In our approach to the weak radiative decays, we  use eqs. (5)(6)
with formfactors from the quark model 
as determined previously 
for the $(V-A)$ current induced transitions.
To calculate the decay rates for the various processes,
we proceed as follows.
For the $\displaystyle\frac{1}{2}\to\displaystyle\frac{1}{2}\gamma$ transitions,
we use (5) with the formfactors of Ref. [11] with monopole behavior
and pole masses $M_V=5.42 $GeV and $M_A=5.85$GeV, 
which gives
\begin{equation}
F_1(q^2=0)=0.059 (0.11), F_2(q^2=0)=- 0.025 (- 0.019),
\end{equation}
where the two given values are  for the vector (axial) currents.
Using now (7) and (1) with $V_{tb}=1$,  $V_{ts}=0.04$,
this leads to
\begin{equation}
\Gamma(\Lambda_b\rightarrow\Lambda\gamma)=1.45\times10^{-15}{\rm MeV};
~~\Gamma(\Xi_b\rightarrow\Xi\gamma)=2.18\times10^{-15}{\rm MeV}.
\end{equation}
If we use a lifetime of $\tau(\Lambda_b)=1.04$ps\cite{aleph},
which has an experimental uncertainty of about $20\%$,
we arrive at a predicted branching ratio 
for $\Lambda_b\rightarrow\Lambda\gamma$
\begin{equation}
BR(\Lambda_b\rightarrow\Lambda\gamma)\simeq 2.3\times 10^{-6}.
\end{equation}
A similar figure would be obtained for $\Xi_b\rightarrow\Xi\gamma$ decays,
except that no measurement exists at present for the  $\Xi_b$ lifetime.
Our result in (9) is smaller than the $10^{-5}$ order of magnitude
obtained in \cite{10}, 
though quite close to the recent result of Ref. \cite{11}.

A similar approach is used for the decay $\Omega_b\to\Omega\gamma$.
In this case,
using the quark model of Ref.\cite{13}, we obtain
\begin{equation}\begin{array}{rcl}
C_1(q^2)&=&\displaystyle\frac{(m_{\Omega_b}+m_{\Omega})^2-q^2}
{4m_{\Omega_b}m_{\Omega}}H(q^2),\\[5mm]
C_3(q^2)&=&-\displaystyle\frac{1}{2}H(q^2),\\[5mm]
C_2(q^2)&=&C_4(q^2)=C_5(q^2)=C_6(q^2)=0,\\[5mm]
H(q^2)&=&\sqrt{3}
\displaystyle\frac{1-(m_{\Omega_b}-m_{\Omega})^2
/m_{pole}^2}
{1-q^2/m_{pole}^2},
\end{array}
\label{hu-kor}
\end{equation}
where again a monopole behavior, 
but with same pole mass of $5.42$GeV has been used for all the formfactors.
Taking a mass for $\Omega_b$ of $6.08$GeV, we get
\begin{equation}
\Gamma(\Omega_b\rightarrow\Omega\gamma)=1.63\times 10^{-13} {\rm MeV}.
\label{gammaome}
\end{equation}
It is interesting to remark that  transitions involving
$\Omega_c\to \Omega$ are similarly larger than those for
$\Lambda_c\to\Lambda$ and $\Xi_c\to\Xi$
by one or two orders of magnitude in the decay rates\cite{13,14,19}.
These results are apparently a result of the strong ovelaps in the
light flavour wavefunctions between the initial and final states in
$\Omega_{b,c}\to\Omega$ transitions.
Additional support for this picture is the fact that 
the lifetime of $\Omega_{c}$,
recently measured to be $0.055\times 10^{-12}$sec\cite{20},
is considerably shorter than those of other weakly
decaying charmed hadrons,
which can also be understood qualitatively in this way\cite{21}.
The branching ratio for $\Omega_b\rightarrow\Omega\gamma$ is then
\begin{equation}
{\rm Br}(\Omega_b\rightarrow\Omega\gamma)=
1.3\times10^{-4}[
\displaystyle\frac{\tau(\Omega_b)}{0.5\times 10^{-12}{\rm sec}}],
\label{brome}
\end{equation}
where we have normalized the $\Omega_b$ lifetime to $0.5\times 10^{-12}$sec.
The branching ratio in (\ref{brome}) is larger by two orders of magnitude
than that of $\Lambda_b\to\Lambda\gamma$ in (10).
Thus, from the present calculation, the $\Omega_b\rightarrow\Omega\gamma$
decay mode is appearing as the most attractive baryonic exclusive 
channel for an alternative investigation
of the $b\to s\gamma$ electroweak penguin.

We conclude by a discussion on additional
advantages of studying the weak radiative decay
of the beauty baryons.
Among these processes, $\Xi_b^-\rightarrow\Xi^-\gamma$ and
$\Omega_b\rightarrow\Omega\gamma$ are the cleanest
on the theoretical side.
Neither of these processes involves
 the W-exchange or the W-annihilation
diagrams, which give additional  contributions
in the weak radiative decays 
of  other $b$-hadrons.
Also, the long distance contributions in this $b$-sector
are small and under control\cite{22}.
Thus ,  the measurements of $\Xi_b^-\rightarrow\Xi^-\gamma$ and of
$\Omega_b\rightarrow\Omega\gamma$  in future experiments, together
with  their improved theoretical estimations, will be helpful
to  obtain the needed insight on  long distance contribution in
weak radiative decays of the
other beauty hadrons, as well as
on the question of $W$-exchange contributions in those decays.

\vskip 0.8cm

This research  is supported in part by Grant 5421-3-96
from the Ministry of Science and the Arts of Israel. 
The research of P.S. has also been supported 
in part by the Fund for Promotion of Research
at the Technion.

\end{document}